\documentstyle[aps,prl,twocolumn,epsf]{revtex}
\begin{document}

\draft
\title{Spin and interaction effects in quantum dots: a Hartree-Fock-Koopmans 
approach} 

\author{Y. Alhassid and S. Malhotra}

\address{Center for Theoretical Physics, Sloane Physics Laboratory, Yale 
University,  New Haven, Connecticut 06520, USA} 

\date {submitted February 27, 2002}

\maketitle

\begin{abstract}
We use a Hartree-Fock-Koopmans approach to study spin and interaction effects 
in  a diffusive or chaotic quantum dot. In particular, we derive the 
statistics of the spacings between successive Coulomb-blockade peaks. We 
include  fluctuations of the  matrix elements of the two-body 
screened interaction,  surface-charge potential, and confining potential to 
leading order  in the inverse Thouless conductance. The calculated
peak-spacing distribution is compared with  experimental results. 
\end{abstract}


\vspace{3 mm}

\narrowtext

The traditional description of Coulomb blockade in quantum dots has been the 
constant-interaction  (CI) model, in which the electrons occupy single-particle 
levels  in a confining potential,  and the interaction is taken as a constant 
charging  energy. In dots with chaotic dynamics, the fluctuations of the 
single-particle  levels and wave functions can be described by random-matrix 
theory  (RMT).  The CI plus RMT model was successful in describing some of the 
observed  statistical properties of such dots, e.g., the conductance 
peak-height  distribution \cite{Alhassid00,JSA92}. However, several 
experiments  have demonstrated that other statistics, such as the distribution 
of  the spacing between Coulomb-blockade peaks, are affected by 
electron-electron  interactions \cite{Sivan96,Simmel97,Patel98,Luscher01}.

    At low temperatures the peak spacing is given by the second-order 
difference  of the ground-state energy versus the number of electrons. In the 
CI  model and for spin-degenerate  levels, the peak-spacing distribution is 
bimodal,  i.e., a superposition of a $\delta$ function and a (shifted) 
Wigner-Dyson  distribution.  However, none of the measured distributions are 
bimodal  and they all deviate from Wigner-Dyson statistics 
\cite{Sivan96,Simmel97,Patel98,Luscher01}.  

   In the spinless case, exact diagonalization for a small number of  electrons\cite{Sivan96}, Hartree-Fock (HF) \cite{HF} and density functional 
\cite{Hirose01}  calculations, as well as a random interaction matrix model 
\cite{Alhassid00a}  explained the deviation from Wigner-Dyson statistics as an 
interaction  effect. Spin degrees of freedom were included using exact 
diagonalization  for dots with a small number of electrons and small values of 
the  Thouless conductance $g$ \cite{Berkovits98}. Spin effects 
were  studied in the limit $g \to \infty$ \cite{Oreg01,Ullmo01,Usaj01} using 
the  so-called universal Hamiltonian \cite{Kurland00,Aleiner02}.  Interaction 
effects (in the presence of a magnetic field) were included using a Strutinsky 
approach  for quantum dots \cite{Ullmo01}. The resulting distributions showed 
qualitative  differences when compared with experiments. Temperature 
effects  were shown to be important at temperatures as low as $T \sim 0.1 - 
0.2\Delta$  \cite{Usaj01} ($\Delta$ is the mean spacing between 
spin-degenerate levels). The spacing statistics were also studied in a spin 
density  functional theory for dots with $\sim 10$ electrons \cite{Hirose02}.

 Here we develop a theory that includes spin and interaction effects and is 
based  on a HF-Koopmans approach. The theory is generic and does not require 
the  actual solution of the HF equations.  Rather than using the 
non-interacting  basis, we choose as a reference state the $n$-electron dot 
with  spin $S=0$ ($n$ even) and work in its HF basis. We then consider the 
addition  energies and the energy differences between various spin 
configurations  in Koopmans' limit where the single-particle wave functions do 
not  change \cite{Koopmans34}.  Previously, Koopmans' approach was discussed 
for  dots with {\em spinless} electrons \cite{HF} and was used to study 
spectral  scrambling \cite{Alhassid01}.  Here we derive the generic statistics 
of  the spin and peak spacings, assuming that the HF levels of the reference 
state  satisfy statistics typical of a diffusive or chaotic dot. The theory is 
valid  for large $g$, and for $g \to \infty$ it reduces to the universal 
Hamiltonian  \cite{Kurland00}. For finite $g$, we include fluctuations of the 
diagonal  matrix elements to leading order in $1/g$. Compared with the approach of Ref.~\cite{Ullmo01},  our theory requires the statistics of only a few levels 
around  the Fermi energy, and some of its results are qualitatively different.

The Hamiltonian of the quantum dot in the ``disorder'' basis
 $|i \; \sigma \rangle  = a_{i\sigma}^\dagger |0\rangle$  ($i$ denotes a 
spatial  orbit and $\sigma=\pm 1$ is the spin variable) is given by 
\begin{eqnarray}\label{Hamiltonian}
H = \sum\limits_{i \sigma} \epsilon^{(0)}_{i} a^\dagger_{i \sigma} a_{i 
\sigma}  +{1\over 2} \sum_{ijkl\atop \sigma \sigma'}
v_{ij;kl}a^\dagger_{i \sigma} a^\dagger_{j \sigma'} a_{l \sigma'} a_{k 
\sigma} 
\;,
\end{eqnarray}
where $\epsilon^{(0)}_i$ are the single-particle energies and $v_{ij;kl}$ are 
the  (spin-independent) matrix elements of the Coulomb interaction.
The Hamiltonian (\ref{Hamiltonian}) can be solved in the HF approximation. 
For  each value of the spin projection $S_z$, we use  Slater determinants with 
$n_+$  ($n_-$) spin up (down) orbitals. Usually, the HF single-particle 
energies  $\epsilon^{(n)}_{\alpha\sigma}$ and orbitals $\phi_{\alpha \sigma}$ 
depend  on the spin $\sigma$. However for even $n$, the HF equations have a 
solution  where $\epsilon^{(n)}_\alpha$  and $\phi_\alpha$ are independent of 
$\sigma$,  and the lowest $n/2$ levels are doubly occupied. Such a Slater 
determinant  has good $S=0$. We choose this solution as our reference state, 
and  work in its HF basis $|\alpha\rangle$. This $S=0$ state is an eigenstate 
of  the following diagonal many-particle Hamiltonian  
\begin{equation}\label{HF-spin}
H_{\rm d} = \sum_{\alpha \sigma} \epsilon^{(0)}_{\alpha \alpha} \hat 
n_{\alpha\sigma}  +  \frac{1}{2} \sum_{\alpha\beta \atop \sigma} 
v^A_{\alpha\beta}  \hat n_{\alpha \sigma} \hat n_{\beta \sigma} +  
\sum_{\alpha\beta}  v_{\alpha\beta} \hat n_{\alpha +} \hat n_{\beta -}
\;,
\end{equation}
 where $\hat n_{\alpha \sigma}$ is the number operator of the state $\alpha 
\sigma$,  and $v_{\alpha\beta}\equiv v_{\alpha\beta;\alpha\beta}$, $v^{\rm 
ex}_{\alpha\beta}\equiv  v_{\alpha\beta;\beta\alpha}$, and 
$v^A_{\alpha\beta}\equiv  v_{\alpha\beta} - v^{\rm ex}_{\alpha\beta} $  are 
diagonal,  exchange and antisymmetrized matrix elements, respectively. 

 The Hamiltonian (\ref{HF-spin}) also has eigenstates with $S \neq 0$. We 
label  by $\alpha=0$ the last occupied level of the $S=0$ state. The lowest 
energy state  for each spin $S$ is obtained by promoting $S$ spin down 
electrons from 
$\alpha  =0,\ldots, -(S-1)$ to spin up electrons in $\alpha=1, \ldots ,S$.  
The  resulting Slater determinant  describes a maximal spin projection state 
$S_z=S$  and has good spin $S$. Using (\ref{HF-spin}), the energy difference 
$\delta  E_{n}(S)\equiv E_n(S) - E_n(S=0)$ between the lowest states with spin 
$S$  and spin $S=0$ can be written in terms of $\epsilon^{(n)}_{\alpha}$ and a 
few  matrix elements. For example 
\begin{eqnarray}\label{even-n}
\delta E_n(S=1) &= & (\epsilon^{(n)}_{1} -\epsilon^{(n)}_{0})  - v_{10} \;.
\end{eqnarray}
The spin $S_n$ of the ground state of the $n$-electron dot is determined by 
minimizing  $\delta E_n(S)$. 

 Similarly, the ground-state spin $S_{n+1}$ of the dot with $n+1$ electrons 
can  be determined from the energy differences $\delta E_{n+1}(S) \equiv 
E_{n+1}(S)  - E_{n+1}(S=1/2)$ for half-integer $S$. Assuming Koopmans' limit, 
we  can express $\delta E_{n+1}(S)$ in terms of the HF levels and matrix 
elements  of the reference state $(n,S=0)$.  For example
\begin{eqnarray}\label{odd-n+1}
\delta E_{n+1}(S =3/2)  = 
 (\epsilon^{(n)}_{2} -\epsilon^{(n)}_{0})  - v_{10} - v_{20} +v^A_{21}\;.
\end{eqnarray}

The addition energy $\mu_{n+1} \equiv E_{\rm gs}(n+1) - E_{\rm gs}(n)$ is 
\begin{equation}\label{mu-n+1}
\mu_{n+1} = \mu_{n+1}(0 \to 1/2) + \delta E_{n+1}(S_{n+1}) -  \delta 
E_{n}(S_n)  \;,
\end{equation}
where in Koopmans' limit  $\mu_{n+1}(0 \to 1/2)= \epsilon^{(n)}_{1}$.
 
 The spacing $\Delta_2$ between successive peaks is given by the difference 
in  addition energies. We have to distinguish between even-odd-even (``odd'') 
and  odd-even-odd (``even'') transitions (in particle number).  We consider 
here  the odd transition $n\to n+1 \to n+2$ ($n$ even), for which 
$\Delta_2=\mu_{n+2}  -\mu_{n+1}$ (similar results can be derived for the even 
transition  \cite{Alhassid02}). $\mu_{n+1}$ is given by (\ref{mu-n+1}) and 
$\mu_{n+2}$  is calculated from 
$\mu_{n+2}  =  \mu_{n+2}(1/2 \to 0) + \delta E_{n+2}(S_{n+2}) -  \delta 
E_{n+1}(S_{n+1})$  with  $\mu_{n+2}(1/2 \to 0)  = \epsilon^{(n)}_1 + v_{11}$. 
$\delta  E_{n+1}(S)$ is given by, e.g., Eq. (\ref{odd-n+1}), while $\delta 
E_{n+2}(S)$  is calculated from, e.g.,
\begin{eqnarray}\label{even-n+2}
\delta E_{n+2}(S=1) =  (\epsilon^{(n)}_{2} -\epsilon^{(n)}_{1})  - v_{11} + 
v^A_{21}  \;.
\end{eqnarray}

 To describe the statistics of $\Delta_2$ it is necessary to model the 
fluctuations  of the HF levels and wave functions of the $(n,S=0)$ dot. The 
spectrum  is assumed to follow RMT within $g$ levels around the Fermi energy, 
and  the matrix elements are uncorrelated from the single-particle spectrum.  
An  exception is the gap $\epsilon_1^{(n)} - \epsilon_0^{(n)}$.  We find its 
statistics  by comparing the single-particle spectrum of the $(n+2)$-electron 
dot  with the spectrum of the $n$-electron dot (both in their $S=0$ 
configuration).  We have the relation $\epsilon_1^{(n)} - \epsilon_0^{(n)} = 
\epsilon_1^{(n+2)}  - \epsilon_0^{(n+2)} +v_{01}^A + v_{01} - v_{11}$, in 
which  
  the levels $\epsilon_0^{(n+2)}$ and $\epsilon_1^{(n+2)}$ are both doubly 
occupied,  and thus their spacing should follow Wigner-Dyson statistics. The 
gap  distribution is then a convolution of a Wigner-Dyson distribution with a 
Gaussian  describing the distribution of $v_{01}^A + v_{01} - v_{11}$ (see 
below). 

We apply our HF-Koopmans approach in a restricted single-particle space 
of $\sim g$ levels around the Fermi energy. The long-range bare Coulomb 
interaction should then be replaced by an effective interaction. In the limit
$r_s<<1$ we employ an effective RPA interaction calculated by excluding 
particle-hole transitions within the above strip of 
$\sim g$ levels \cite{Aleiner02}.  This effective interaction is 
$v(\bbox r_1,\bbox r_2) = {e^2 / C} + v_\kappa(\bbox r_1,\bbox r_2) + V(\bbox 
r_1)  + V(\bbox r_2)$,
where $v_\kappa(\bbox r_1,\bbox r_2)$ is a two-body screened 
interaction in the dot, and $V(\bbox r)$ is a one-body potential  
generated by the accumulation  of surface charge in the finite 
dot \cite{Blanter97}.  

We decompose the interaction in (\ref{HF-spin}) into its average and 
fluctuating  parts, and denote by $U_d =  \bar v_{\alpha\beta}$ and $J_s = 
\bar  v^{\rm ex}_{\alpha\beta}$ ($\alpha \neq\beta$) the average values of the 
direct  and exchange interaction, respectively.  The average interaction is 
invariant  under a change of the single-particle basis when 
$\bar v_{\alpha\alpha}  = U_d + J_s$
and can be written in terms of the number operator  $\hat n$ and total spin 
$\bbox  S$ \cite{Cooper}
\begin{eqnarray}\label{V-average}
 \bar V = {1\over 2}\left(U_d - {J_s\over 2}\right) \hat n^2 
- \left({U_d \over 2} -J_s\right)\hat n - J_s \bbox S^2 \;. 
\end{eqnarray}

   The average interaction is determined by the spatial correlations of the 
single-particle  wave functions  \cite{Blanter97,Mirlin00}. 
To leading order in $1/g$, 
$J_s = (\Delta/2)(1+ 2\beta^{-1} {b_1 / g})$ \cite{delta}, where $b_1$ is a
geometry-dependent coefficient determined from $b_1/g=\int d 
\bbox  r \Pi(\bbox r, \bbox r)/{\cal A}$. Here $\Pi(\bbox r_1, \bbox r_2)$ 
is the diffuson propagator and ${\cal A}$ is the area of the dot. 

\begin{figure}
\epsfxsize= 8.2 cm 
\centerline{\epsffile{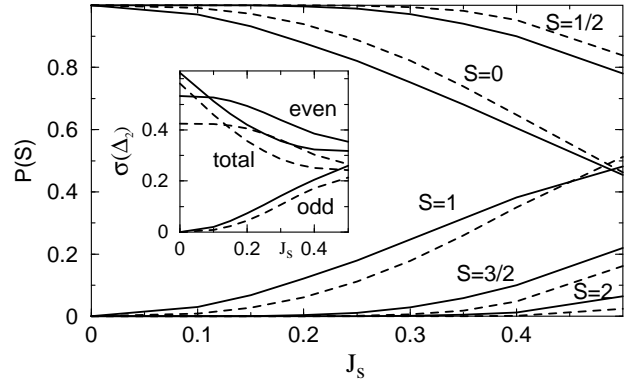}}
\vspace{2 mm}

 \caption {Statistics in the absence of fluctuations of the interaction 
matrix  elements. The probabilities of various spin values $S$ are shown 
versus   $J_s$. The solid (dashed) lines describe the orthogonal (unitary) 
symmetry.  Inset: the standard deviation of the peak spacing 
$\sigma(\Delta_2)$  versus $J_s$ for the even, odd and combined (``total'') 
transitions.  $J_s$ and $\sigma(\Delta_2)$ are measured in units of $\Delta$.}
\label{fig1}
\end{figure}

Koopmans' limit is exact when the fluctuations of the matrix elements are 
ignored,  and our calculations of $\Delta_2$ reduce to those in Ref. 
\cite{Oreg01},  i.e., they can be derived directly from an Hamiltonian that 
consists  of a random one-body part plus an average interaction 
(\ref{V-average}).  In the limit $g \to \infty$ this Hamiltonian is just the 
universal  Hamiltonian with $J_s=\Delta/2$ \cite{Kurland00,Aleiner02}. A 
better  estimate of $J_s$ can be obtained in RPA \cite{Oreg01}; it increases 
monotonically  with $r_s$ but remains below $\Delta/2$.

  In the simple limit (\ref{V-average}), the spin and spacing distributions 
are  determined by a single parameter $J_s/\Delta$ \cite{Oreg01}.  This limit 
is  demonstrated in Fig. \ref{fig1} where we show the probabilities of various 
spin  values versus $J_s$. The inset shows the standard deviation of 
$\Delta_2$   versus $J_s$ for the even and odd transitions as well as the 
combined  one (``total''). 

Next we discuss the fluctuations of the interaction matrix elements 
\cite{Blanter97,Altshuler97}, which  
are  approximately Gaussian variables. We discuss separately the bulk screened 
interaction  and surface-charge potential. Using the diagrammatic approach for 
the  two-body screened interaction, we have, for $r_s\ll 1$ and to leading 
order  in $1/g$ 
\begin{eqnarray}\label{2-body}
{\rm GOE}: \sigma(v_{\alpha \beta})& = & 2\sigma_2;\;\;\sigma(v^{\rm 
ex}_{\alpha  \beta})  = \sqrt{2}\sigma_2; \;\; \sigma(v_{\alpha \alpha}) =  
2\sqrt{2}\sigma_2  \nonumber \\
{\rm GUE}: \sigma(v_{\alpha \beta})& = & \sigma_2;\;\;\sigma(v^{\rm 
ex}_{\alpha  \beta})  =  \sigma_2; \;\; \sigma(v_{\alpha \alpha}) =  
\sqrt{2}\sigma_2  \;.
\end{eqnarray}
Different matrix elements (including the direct $v_{\alpha \beta}$ and 
exchange  $v^{\rm ex}_{\alpha \beta}$) are uncorrelated. We note that the 
coefficients of $\sigma_1$ and $\sigma_2$ in Eqs. (\ref{2-body}) are different from those obtained for a  zero-range 
interaction  \cite{Mirlin}.
The parameter $\sigma_2$ is given by 
\begin{equation}\label{sigma_2}
\sigma_2= \left[ {\cal A}^{-2} \int d\bbox r_1 \int d\bbox r_2 \Pi^2(\bbox 
r_1,\bbox  r_2)\right]^{1/2} =  c_2 {\Delta / g} \;,
\end{equation}
  where $c_2$ is a geometry-dependent coefficient. For a disk of radius $R$ 
and  boundary conditions of vanishing normal derivative, we find 
\cite{Alhassid02}  $c_2= 2 [\sum_{l,m} x_{l,m}^{-4}]^{1/2} \approx 0.67$ where 
$x_{l,m}$  are the zeros of $J_l^\prime(x)$ ($J_l$ is the Bessel function of 
order  $l$) and $g \Delta= 2\pi \hbar D/R^2$ ($D$ being the diffusion 
constant).   Since only a few matrix elements contribute to the peak spacing, 
the  contribution of the two-body screened interaction is  parametrically of 
the  order $\Delta/g$, unlike the  $\Delta/\sqrt{g}$ dependence found in Ref. 
\cite{Ullmo01}.   

The surface-charge contribution to an interaction matrix element is 
$v_{\alpha  \beta}=V_\alpha + V_\beta$, where 
$V_\alpha \equiv \int |\psi_\alpha 
(\bbox  r)|^2 V(\bbox r)$ is a diagonal matrix element of the surface-charge 
potential.  We have
\begin{eqnarray}\label{1-body}
\sigma(V_\alpha) =  (2/ \beta)^{1/2} \sigma_1 \;; \;\; \overline{V_\alpha 
V_\beta}  - \overline{V_\alpha}\; \overline{V_\beta}  \approx 0
\;,
\end{eqnarray}
where 
\begin{eqnarray}\label{sigma_1}
\sigma_1 & = & \left[ {\cal A}^{-2} \int d\bbox r_1 \int d\bbox r_2  
V(\bbox  r_1)\Pi(\bbox r_1,\bbox r_2) V(\bbox r_2) \right]^{1/2}  \nonumber \\
& =  & c_1 {\Delta  / \sqrt{g}} \;.
\end{eqnarray}
For an isolated two-dimensional (2D) circular disk of radius $R$, the 
surface-charge  potential can be approximated by 
$V(\bbox r) = -(e^2/2\kappa \epsilon
R)  (R^2-r^2)^{-1/2}$, where $\kappa=2\pi e^2 \nu/\epsilon$ is the inverse 
screening  length in 2D 
and $\epsilon$ is the dielectric constant  \cite{Blanter97}.
 We  then find $c_1=  2^{-1/2} [\sum_{m\neq 0} \sin^2 x_{0,m}/(x_{0,m}^4 
J_0^2(x_{0,m}))]^{1/2}  \approx 0.087$ \cite{Alhassid02}.

The above results can be generalized to a ballistic dot, using the ballistic 
supersymmetric  $\sigma$ model obtained when a weak disorder with finite 
correlation  length is added \cite{Gornyi01}. In particular, if the 
Lyapunov  length of this smooth disorder is smaller than the dot's size,  
relations similar to Eqs. 
(\ref{2-body}) and (\ref{1-body}) can be derived but with the ballistic propagator $\Pi_B$ 
replacing the diffusive propagator $\Pi$ in Eqs. (\ref{sigma_2}) and (\ref{sigma_1}). For a circular dot we define the ballistic 
Thouless conductance from the inverse time it takes to 
cross  the diameter $2R$ of the dot, leading to $g=\pi k_F R/2 =\pi 
(n/2)^{1/2}$ ($n$ is the number of electrons in the dot).  In the ballistic 
case
\begin{equation}
\sigma_2= c_2 \Delta [\ln(c^\prime_2 g)]^{1/2}/g
\;,
\end{equation} 
where $c_2=\sqrt{3}/2$ is a geometry-independent constant.  The direct 
propagation  between $\bbox r_1$ and $\bbox r_2$ contributes to 
$\Pi_B({\bbox r}_1,  {\bbox r}_2)$ a term $1/(\pi k_F |\bbox r_1 -
 \bbox r_2|)$ which at shorter distances should be replaced by its quantum 
counterpart  
$J_0^2(k_F|{\bbox r}_1 -{\bbox r}_2|)$ \cite{correlator}. The corresponding 
contribution needs to be renormalized such that its integral over $\bbox r_1$
 (or $\bbox r_2$) vanishes \cite{Blanter01}.  
 The remaining part of $\Pi_B$ involves single or multiple scattering from the
 boundaries  and calculating it requires knowledge of the semiclassical dynamics. It can be
 calculated  analytically for a circular billiard with diffusive boundary 
scattering \cite{Blanter01}.   Using this model, we estimate  
$c^\prime_2=0.81$  \cite{Alhassid02}. A similar estimate of (\ref{sigma_1}) 
gives  $c_1= 0.123$.

\begin{figure}
\epsfxsize= 8.2 cm 
\centerline{\epsffile{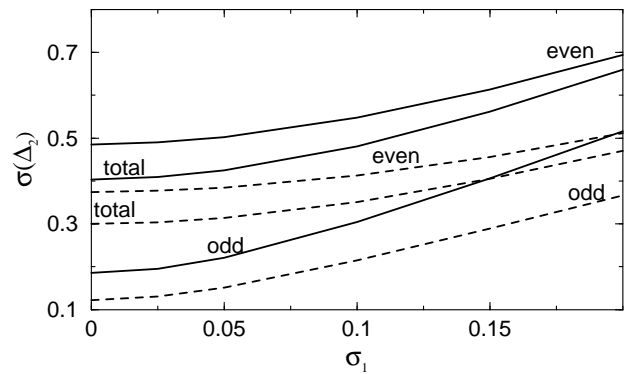}}
\vspace{2 mm}

  \caption {The standard deviation $\sigma(\Delta_2)$ versus the standard 
deviation  $\sigma_1$ of the surface-charge potential for $J_s=0.3  \Delta$ 
and   $\sigma_2=0.05 \Delta$. The solid and dashed lines describe the 
orthogonal  and unitary symmetries, respectively.}
\label{fig2}
\end{figure}

We have studied the effects of fluctuations of the interaction matrix 
elements  on the peak-spacing distribution. In Fig. \ref{fig2} we show the 
standard  deviation $\sigma(\Delta_2)$ versus  $\sigma_1$ ($J_s=0.3 \Delta$ 
and  $\sigma_2=0.05 \Delta$) for the orthogonal (solid lines) and unitary 
(dashed  lines) symmetries. $\sigma(\Delta_2)$ increases with $\sigma_1$ and 
it  does so faster in the odd case. 

\begin{figure}
\epsfxsize= 8.2 cm 
\centerline{\epsffile{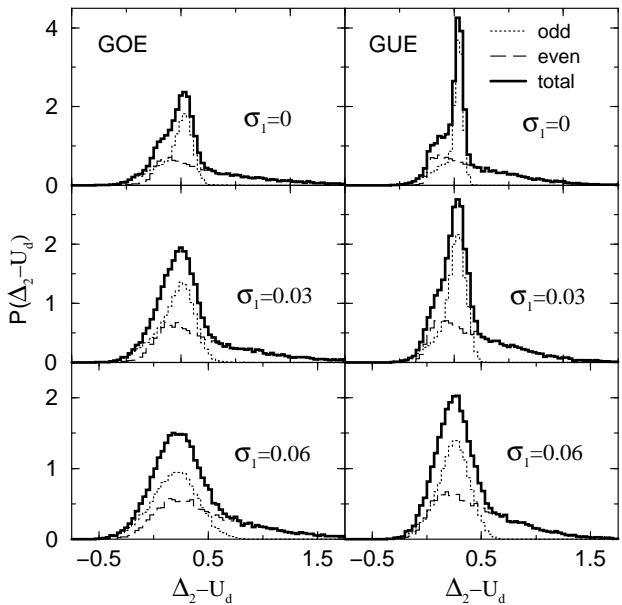}}
\vspace{2 mm}

\caption {Peak-spacing distributions $P(\Delta_2)$  for the orthogonal (left) 
and  unitary (right) symmetries and for $\sigma_1=0, 0.03 \Delta$ and $0.06 
\Delta$.  In all cases $J_s=0.3\Delta$ and $\sigma_2=0.025\Delta$. The 
bimodality  is lost already for $\sigma_1 = 0.03\Delta$ but the distributions 
remain  asymmetric.}
\label{fig3}
\end{figure}

Fig. \ref{fig3} shows typical peak-spacing distributions for both the 
orthogonal  (left) and unitary (right) symmetries for $\sigma_2 =0.025 \Delta$ 
and  $\sigma_1 =0$, $0.03 \Delta $ and $0.06 \Delta $. Signatures of the 
bimodality  can still be observed for $\sigma_1=0$ but they disappear at 
$\sigma_1=0.03  \Delta $. Nevertheless, the distributions remain asymmetric 
(more  so in the unitary case). 

  An additional contribution to the fluctuations of $\Delta_2$ arises from 
the  variation of the gate voltage between  peaks. In general, the change of 
the  gate voltage between two peaks leads to a spatially non-uniform change 
$U(\bbox  r) = - V(\bbox r) + {\tilde V}(\bbox r)$ in the confining potential, 
where  $\tilde V$ originates in the mutual dot-gate capacitance 
\cite{Aleiner02}.   This leads to scrambling of the HF levels between peaks. 
For  example, let us consider the odd  transition. As the gate voltage changes 
between  $V_g^{n+1}$ and $V_g^{n+2}$,  the reference  HF levels 
$\epsilon_\alpha^{(n)}$  change by $\delta \epsilon_\alpha^{(n)} \approx 
U_{\alpha}  = \int d \bbox r |\psi_\alpha(\bbox r)|^2 U(\bbox r)$.  We 
therefore  substitute $\epsilon_\alpha^{(n)} \to \epsilon_\alpha^{(n)}
 + U_\alpha$ in the 
calculation  of $\mu_{n+2}$ and in Eq. (\ref{even-n+2}) ($\mu_{n+1}$ is 
unchanged  and calculated from Eq. (\ref{mu-n+1})). This level  scrambling can 
also  lead to spin rearrangement in the dot. The ground-state spin of the 
$n+1$-electron  dot at gate voltage $V_g^{n+2}$ (just below the transition) is 
found  from (\ref{even-n+2}) (and its generalization to higher values of $S$) 
after  the substitution $\epsilon_\alpha^{(n)} \to \epsilon_\alpha^{(n)}
 + U_\alpha$.

  The fluctuation properties of $U_\alpha$ are similar to those of 
$V_\alpha$  in Eqs. (\ref{1-body}), except that $V(\bbox r)$ is replaced by 
$U(\bbox  r)$ in Eq. (\ref{sigma_1}). 

\begin{figure}
\vspace{3 mm}
\epsfxsize= 7 cm 
\centerline{\epsffile{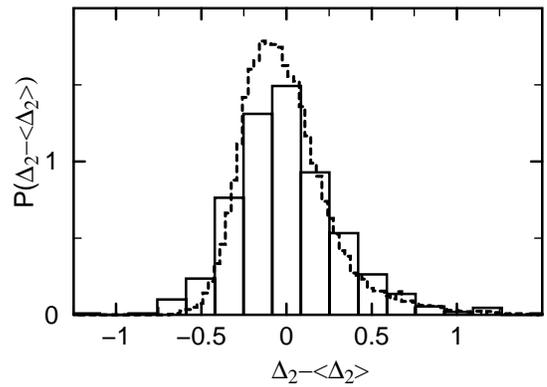}}
\vspace{2 mm}

 \caption {Calculated peak-spacing distribution at $T=0.22 \Delta$ for the 
unitary symmetry (dashed histograms) and $J_s=0.28\Delta$,  
$\sigma_1= 0.02 \Delta$ 
and $\sigma_2= 0.04 \Delta$ is compared  with the experimental distribution of
 Ref. \protect\cite{Patel98} 
(solid  histograms)}
 \label{fig4}
\end{figure}

  We now compare our theory with the experimental results of Ref. \cite{Patel98} 
at the lowest measured temperature of $T=0.22 \Delta$.  At this temperature 
it is necessary to include the effect of excited states and in particular the
 contribution from both lowest $S=0$ and $S=1$ states \cite{Alhassid02}. 
We model the gate-voltage scrambling by a potential $\tilde V$ whose matrix 
elements are uncorrelated from the matrix elements of $V$ and have the same 
variance. It 
is difficult to estimate $\sigma_1$ and $\sigma_2$ for the 
ballistic  dot used in the experiment.  The simple estimates based on a 
billiard  with diffusive surface scattering (see the second paragraph after 
Eq. 
(\ref{sigma_1}))  for $n \approx 340$ electrons (i.e., $g \approx 41$) give 
$\sigma_1=  0.02 \Delta$ and $\sigma_2= 0.04 \Delta$.
Fig. \ref{fig4} compares the experimental distribution of Ref. \cite{Patel98} 
in the presence of a magnetic field (solid histograms) with the corresponding
 theoretical distribution (dashed
histograms) that includes an experimental noise of $0.1 \Delta$.  The
 theoretical distribution describes rather well the asymmetry of the 
experimental distribution and its width $0.27\,\Delta$ is slightly below the 
experimental width of $(0.29 \pm 0.03) \Delta$.  In chaotic billiards 
our estimates of $\sigma_1$ and $\sigma_2$ can be enhanced by up to a factor 
of $2$ and lead to better agreement with the data.   

In conclusion, we have developed a HF-Koopmans approach to study spin and 
interaction  effects in diffusive or chaotic quantum dots.  In particular we 
have  studied the dependence of the peak-spacing distribution on the 
fluctuations  of the interaction matrix elements to leading order in the 
inverse  Thouless conductance. We find good agreement with the lowest 
temperature data of Ref.\cite{Patel98}.

 This work was supported in part by the U.S. DOE grant No.\ 
DE-FG-0291-ER-40608.    We acknowledge useful discussions with H.U. Baranger, Y. Gefen, A. Polkovnikov and G. Usaj,  and in particular with A.D. Mirlin.


\begin{references}

\bibitem{Alhassid00} Y. Alhassid, Rev. Mod. Phys. {\bf 72}, 895 (2000).

\bibitem{JSA92} R.A. Jalabert, A.D. Stone, and Y. Alhassid, Phys. Rev. Lett. 
{\bf  68}, 3468 (1992).

\bibitem{Sivan96} U. Sivan {\em et al.}, Phys. Rev. Lett. {\bf 77}, 1123 
(1996). 

\bibitem{Simmel97}  F. Simmel, T. Heinzel and D.A. Wharam,
Europhys. Lett. {\bf  38}, 123 (1997).

\bibitem{Patel98}  S.R. Patel {\em et al.}, Phys. Rev. Lett.
{\bf 80}, 4522 (1998).

\bibitem{Luscher01} S. L\"{u}scher {\em et al.}, Phys. Rev. Lett. {\bf 86}, 
2114 (2001).

\bibitem{HF} S. Levit and D. Orgad, Phys. Rev. B {\bf 60}, 5549 (1999); P.N. 
Walker,  G. Montambaux, and Y. Gefen,
Phys. Rev. B {\bf 60}, 2541 (1999); A. Cohen, K. Richter, and 
R. Berkovits, Phys. Rev. B {\bf 60}, 2536 (1999).

\bibitem{Hirose01} K. Hirose, F. Zhou, and N.S. Wingreen, Phys. Rev. B  {\bf 
63},  075301 (2001).

\bibitem{Alhassid00a} Y. Alhassid, Ph. Jacquod, and A. Wobst, Phys. Rev. B 
{\bf  61}, R 13357 (2000).

\bibitem{Berkovits98} R. Berkovits, Phys. Rev. Lett. {\bf 81},  2128 (1998).

\bibitem{Oreg01} Y. Oreg, P.W. Brouwer, X. Waintal, and B.I. Halperin, 
cond-mat/0109541  (2001).

\bibitem{Ullmo01} D. Ullmo and H.U. Baranger, Phys. Rev. B {\bf 64},  245324 
(2001). 

\bibitem{Usaj01} G. Usaj and H.U. Baranger, Phys. Rev. B {\bf 64}, 201319(R) 
(2001). 

\bibitem{Kurland00} I.L. Kurland, I.L. Aleiner, and B.L. Altshuler, Phys. 
Rev.  B {\bf 62}, 14886 (2000).

\bibitem{Aleiner02} I.L. Aleiner, P.W. Brouwer, and L.I. Glazman, Phys. Rep. 
{\bf  358}, 309 (2002).

\bibitem{Hirose02} K. Hirose and N.S. Wingreen, cond-mat/0202266.

\bibitem{Koopmans34}
T. Koopmans, Physica (Amsterdam) {\bf 1}, 104 (1934).

\bibitem{Alhassid01} Y. Alhassid and Y. Gefen, cond-mat/0101461.

\bibitem{Alhassid02} Y. Alhassid and S. Malhotra, to be published.

\bibitem{Blanter97} Ya. M. Blanter, A.D. Mirlin, and B.A. Muzykantskii, Phys. 
Rev.  Lett. {\bf 78}, 2449 (1997).

\bibitem{Cooper} More generally, an additional Cooper channel interaction 
$J_c=  \bar v_{\alpha\alpha;\beta\beta}$ can contribute to (\ref{V-average}) 
in  the orthogonal case. 

\bibitem{Mirlin00} A.D. Mirlin, Phys. Rep. {\bf 326}, 260 (2000).

\bibitem{delta} Here we approximate the screened interaction by a contact 
interaction  $(\Delta/2){\cal A} \delta(\bbox r_1 - \bbox r_2)$.

\bibitem{Altshuler97} B.L. Altshuler {\em et al.}, Phys. Re. Lett. {\bf 78}, 
2803 (1997).

\bibitem{Mirlin} A.D. Mirlin, private communication.

\bibitem{Gornyi01} I.V. Gornyi and A.D. Mirlin, cond-mat/0107552.

\bibitem{correlator} The proper correlator behaves as 
$J^2_0(k_F |\bbox r_1 - \bbox r_2|)$ at short distances and 
$1/(\pi k_F |\bbox r_1 - \bbox r_2|)$ at large distances.

\bibitem{Blanter01} Ya. M. Blanter, A.D. Mirlin, and B.A. Muzykantskii, Phys. 
Rev.  B {\bf 63} 235315 (2001).

\end{references}
\end{document}